\documentclass[12pt,numreferences]{iopart}

\usepackage{dsfont}
\usepackage{epsfig}
\usepackage{amssymb}
\usepackage{amsfonts}

\begin{document}

\title{Resummation of mass terms in perturbative massless quantum field theory}
\author{Andreas Aste}
\address{Department of Physics and Astronomy, University of Basel,\\
Klingelbergstrasse 82, 4056 Basel, Switzerland}
\eads{andreas.aste@unibas.ch}
\date{\today}

\begin{abstract}
The neutral massless scalar quantum field $\Phi$ in four-dimensional
space-time is considered, which is subject to a simple
bilinear self-interaction. Is is well-known from renormalization
theory that adding a term of the form $-\frac{m^2}{2} \Phi^2$ to the
Lagrangean has the formal effect of shifting the particle mass
from the original zero value to $m$ after resummation
of all two-leg insertions in the Feynman graphs appearing in
the perturbative expansion of the $S$-matrix. However, this resummation
is accompanied by some subtleties if done in a proper
mathematical manner. Although the model seems to be almost trivial,
is shows many interesting features which are useful for the understanding
of the convergence behavior of perturbation theory in general.
Some important facts in connection with the
basic principles of quantum field theory and distribution theory
are highlighted, and a remark is made on possible generalizations of
the distribution spaces used in local quantum field theory. A short
discussion how one can view the spontaneous breakdown of gauge symmetry
in massive gauge theories within a massless framework is presented.
\end{abstract}

%%%%%%%%%%%%%%%%%%%%%%%%%%%%%%%%%%%%%%%%%%%%%%%%%%%%%%%

\section{Introduction}
The traditional starting point of perturbative quantum field theory is
a classical Lagrangean $\mathcal{L}$ which can be decomposed into a free
(solvable) part $\mathcal{L}_0$ and an interacting part $\mathcal{L}_{int}$
which describes the interaction.
These objects get quantized and $S$-matrix elements or Greens functions are
constructed with the help of the Feynman rules.
In this paper, we focus mainly on the extremely simple case of a free scalar massless
field $\Phi$ fulfilling the wave equation
\begin{equation}
\Box \Phi(x) = \partial_\mu \partial^\mu \Phi(x)=0,
\end{equation}
which can be decomposed into a negative and positive frequency
parts according to the representation
($k^0=|\vec{k}|$, $kx=k^\mu x_\mu=k^0 x^0 -\vec{k} \vec{x}$)
\begin{equation}
\Phi(x)=\Phi^-(x)+\Phi^+(x)=(2 \pi)^{-3/2}
\int \frac{d^3k}{\sqrt{2 |\vec{k}|}} \Bigl[ a(\vec{k})
e^{-ikx}+a^\dagger(\vec{k}) e^{ikx} \Bigr]
\end{equation}
with the distributional commutation relations for the creation and
annihilation `operators' \cite{Constantinescu}
\begin{equation}
[a(\vec{k}),a^\dagger(\vec{k}')]=\delta^{(3)}(\vec{k}-\vec{k}'), \quad
[a(\vec{k}),a(\vec{k}')]=[a^\dagger(\vec{k}),a^\dagger(\vec{k}')]=0,
\end{equation}
where the full Lagrangean describing the dynamics of the field
shall be given by
\begin{displaymath}
\mathcal{L}=\mathcal{L}_0 + \mathcal{L}_{int} \quad \mbox{with}
\end{displaymath}
\begin{equation}
\mathcal{L}_0=\frac{1}{2} \partial_\mu \Phi \partial^\mu \Phi, \quad
\mathcal{L}_{int}=-\frac{1}{2} m^2 \Phi^2. \label{Lagrangean}
\end{equation}
Actually, we will avoid the simple way by solving the massive
Klein-Gordon equation $(\Box+m^2)\Phi(x)=0$, which directly leads
to the corresponding free massive scalar field. The aim of this
paper is to highlight in which sense the massive theory comes out
as a perturbative limit of the massless case. This transition,
although seemingly trivial, shows many interesting features which
can be observed also in `more realistic' theories. E.g.,
the positive energy spectrum of the massless theory is continuous, whereas
the energy spectrum of the new massive theory displays a gap
between the vacuum and the lowest one-particle energy state.
An example for an exactly solvable theory which displays such a behavior
is the Schwinger model, i.e. massless quantum electrodynamics in 2+1 space-time
dimensions \cite{Schw1,ASchw}.

The free massless field $\Phi$ is an operator valued distribution in the sense
that it must be smoothly averaged over some space(-time) region
according to the formal expression
\begin{equation}
\Phi(f)=\int \Phi(x) f(x) d^4 x \label{smearing}
\end{equation}
in order to yield an operator which is densely defined in the
corresponding Fock space. E.g., one might assume that $\Phi(x)$ is an operator
applicable to the Fock vacuum $|0 \rangle$. Then a simple calculation using the
relations given above shows that $\Phi(x) |0\rangle$ can not be associated
with a vector in Fock space.
Within the famous Wightman framework \cite{PCT}, $f$
in eq.(\ref{smearing}) is an element of the Schwartz space $\mathcal{S}(\mathds{R}^4)$
or rapidly decreasing functions.
There are several good reasons for the use of the Schwartz space in the Wightman formalism.
The Fourier transform is a linear isomorphism on $\mathcal{S}(\mathds{R}^4)$.
Correspondingly, the Fourier transform acts also as a linear isomorphism on the
dual space of tempered distributions $\mathcal{S}' (\mathds{R}^4)$.
Furthermore, $\mathcal{S}(\mathds{R}^4)$ contains the test functions with compact support
$\mathcal{D}(\mathds{R}^4) \! \subset \! S(\mathds{R}^4)$, which are used to
express the locality and causality properties of field operators.
E.g., the causal property of the scalar field can be characterized by test
functions $f,g \! \in \! \mathcal{D} (\mathds{R}^4)$ via
\begin{equation}
[\Phi(f),\Phi(g)]=0 \, \, \, \forall \, f,g \, \, \, \mbox{with}\, \, \, 
\mbox{supp}(f) \sim \mbox{supp}(g), \label{causality}
\end{equation}
where $\mbox{supp}(f) \sim \mbox{supp}(g)$ denotes the fact that every element
in the support of $f$ is space-like separated with respect to every element in the
support of $g$.

\section{Resummation of the 2-leg insertions}
The theory defined by the Lagrangean eq. (\ref{Lagrangean}) is
ultraviolet super-renormalizable
and contains only diagrams with a very simple topology. Vacuum diagrams have
a polygonial structure, and will play no relevant role in the forthcoming discussion,
i.e. we argue that they lead only to a phase factor in the $S$-matrix and can be
`divided away'.
The $\Phi^2$-term which defines the interaction leads to 2-leg
insertions in the free massless propagator. In renormalization theory, such insertions
are used to switch from the `bare' particle mass to the `physical' mass, but in our case,
we will start from the purely massless theory and investigate the effect of including
an interaction mass term. 

In the literature, a perturbative expansion of the $S$-Matrix according to
\begin{equation}
S={\bf 1}+\sum \limits_{n=1}^{\infty} \frac{(-i)^n}{n !} \int d^4 x_1 ... d^4 x_n
T\{{\cal{H}}_{int}(x_1) {\cal{H}}_{int}(x_2) \cdot ... \cdot {\cal{H}}_{int}(x_n)\},
\label{stoer1}
\end{equation}
is widespread, where $T$ is the time-ordering operator and $\mathcal{H}_{int}$ is the
Hamiltonian interaction density, which is given by $\mathcal{H}_{int}=-\mathcal{L}_{int}$
in our case.
It must be pointed out that the perturbation series eq. (\ref{stoer1}) is formal
and it is difficult to make any statement about the convergence of this series for a
general quantum field theory.
Furthermore, two problems arise in the expansion given above.
First, the time-ordered products
\begin{equation}
T_n(x_1,x_2, ...,x_n)=(-i)^n T\{{\cal{H}}_{int}(x_1) {\cal{H}}_{int}(x_2)
\cdot ... \cdot {\cal{H}}_{int}(x_n)\}
\end{equation}
are usually plagued by ultraviolet divergences when calculated in momentum space
according to the Feynman rules.
Still, these divergences can be removed by
regularization, such that the operator-valued distributions $T_n$ can be viewed as
well-defined, already regularized expressions \cite{GScharf,Aste,AsteSunrise}.
In the present case,
one has $T_1(x)=-\frac{i m^2}{2} : \! \Phi(x)^2 \! :=i : \! \mathcal{L}_{int} \! :$,
such that ultraviolet divergences are absent in the diagrams considered in this paper.
The colons denote normal ordering.
Second, infrared divergences are also present in eq. (\ref{stoer1}).
This is not astonishing, since the $T_n$'s are operator-valued distributions, and
therefore must be smeared out by test functions in $\mathcal{S}(\mathds{R}^{4n})$.
One may therefore introduce
a test function $g(x) \! \in \! \mathcal{S}(\mathds{R}^{4})$ which plays the role of an
'adiabatic switching' and provides a cutoff in the long-range part of the interaction, which
can be considered as a natural infrared regulator \cite{GScharf,eg}.
The correct expression for the infrared regularized $S$-matrix is given by
\begin{equation}
S(g)=
\sum \limits_{n=0}^{\infty} S_n(g)=
{\bf 1}+\sum \limits_{n=1}^{\infty} \frac{1}{n !} \int d^4 x_1 ... d^4 x_n
T_n(x_1,...x_n) g(x_1) \cdot ... \cdot g(x_n),
\end{equation}
and an appropriate adiabatic limit $g \rightarrow 1$ must be performed at the end
of actual calculations in the right quantities (like cross sections)
where this limit exists. This is not one of the standard strategies usually
found in the literature, however, it is the natural one
in view of the mathematical framework used in perturbative quantum field
theory. Performing the adiabatic limit is also necessary to restore the full
Lorentz invariance of the theory.

The Feynman propagator of the free massless scalar quantum field
is given by
\begin{displaymath}
\Delta_F(x)=-i \langle 0 | T (\Phi(x) \Phi(0)) | 0 \rangle
=\int \frac{d^4 k}{(2 \pi)^4} \frac{e^{-ikx}}{k^2+i0}
\end{displaymath}
\begin{equation}
=\frac{i}{4 \pi^2} \frac{1}{x^2-i0} = \frac{i}{4 \pi^2} P \frac{1}{x^2}-
\frac{1}{4 \pi} \delta (x^2),
\end{equation}
where $T$ is the time-ordering operator, $P$ denotes principal value
regularization and $\delta$ is the one-dimensional Dirac distribution depending
on $x^2=x_\mu x^\mu=(x^0)^2-(x^1)^2-(x^2)^2-(x^3)^2=x_0^2-{\vec{x}}^2$.

Usually, the resummation of the particle propagator is performed by a formal
calculation. One observes first that the lowest order non-trivial contribution to the
$S$-matrix is generated by the interaction term
$T_1(x)=-\frac{i m^2}{2} : \! \Phi(x)^2:=i \! : \mathcal{L}_{int} \! :$, which enters the
$S$-matrix at first order in $m^2$ as
\begin{equation}
S_1(g)=-i \frac{m^2}{2}  \int d^4 x_1 : \! \Phi(x_1)^2 \! : g(x_1).
\end{equation}
At second order, one has after Wick ordering of the field operators
\begin{displaymath}
S_2(g)=\frac{4}{2!}\Bigl(\frac{-i m^2}{2} \Bigr)^2 \int d^4 x_1 d^4 x_2
\, i \Delta_F(x_1-x_2) : \! \Phi(x_1) \Phi(x_2) \! : g(x_1) g(x_2)
\end{displaymath}
\begin{displaymath}
 + \, other \, Wick \, contractions
\end{displaymath}
\begin{displaymath}
=-i \frac{m^4}{2} \int d^4 x_1 d^4 x_2 \Delta_F(x_1-x_2)
: \! \Phi(x_1) \Phi(x_2) \! : g(x_1) g(x_2)
\end{displaymath}
\begin{equation}
 + \, other \, Wick \, contractions,
\end{equation}
and for $S_n(g)$ one obtains, taking the permutation symmetry of the $T_n$ into account,
\begin{displaymath}
S_n(g)=-i \frac{m^{2n}}{2} \int d^4 x_1 ... d^4 x_n \Delta_F(x_1-x_2) ... 
\Delta_F(x_{n-1}-x_n)
\end{displaymath}
\begin{equation}
: \! \Phi(x_1) \Phi(x_n) \! : g(x_1) ... g(x_n) + other \, Wick \, contractions.
\label{Sn}
\end{equation}
Assuming that one may perform a adiabatic limits $g(x_i) \rightarrow 1$ in the right
variables, it becomes clear that the particle-particle transition amplitude is
described after a Fourier transform by the resummed Feynman propagator
\begin{equation}
\frac{1}{k^2+i0} + \frac{m^2}{(k^2+i0)^2}+ \frac{m^4}{(k^2+i0)^3}+... = \frac{1}{k^2-m^2+i0}.
\label{summation}
\end{equation}
However, the situation is not as simple as it seems. First, the geometric sequence
eq. (\ref{summation}) does not converge in the sense of tempered distributions in
$\mathcal{S}'(\mathds{R}^4)$. Second, the terms $\sim (k^2+i0)^{-n}$ for $n \ge 2$
are not uniquely defined in a distributional sense. These terms are too singular for
small momenta $k$. This situation can be compared to the case of scalar fields in
$1+1$ space-time dimensions. There, even the zero mass limit of the propagator
$(k^2-m^2+i0)^{-1}$ does not exist, since $1/k^2$ is not a tempered distribution
in two dimensions \cite{Coleman}. There is no free massless scalar field theory in
two dimensions which does not violate the Wightman axiom of positivity
\cite{Positivity,Abdalla} after infrared regularization. 
Third, the geometric series eq. (\ref{summation}) does not converge for arbitrary $k$.
We will investigate these defects in the following section.

\section{Solutions of the n-fold iterated wave equation}
It is obvious that the expressions $\sim (k^2+i0)^{-n}$ for $n \ge 2$ are infrared
divergent for $k=0$ are therefore need to be regularized in a way consistent
with the basic principles of local quantum field theory. If this were not possible,
the theory would be {\emph{infrared non-renormalizable}.

In order to accomplish this task, we consider first
commutation relations for massless scalar fields which define the
positive and negative frequency massless Jordan-Pauli distributions
\begin{equation}
\Delta^\pm(x)=i [\Phi^{\mp}(x) , \, \Phi^{\pm}(0)] =
i \langle 0 |  [\Phi^{\mp}(x) , \, \Phi^{\pm}(0)] | 0 \rangle \, ,
\label{commutator_JP}
\end{equation}
which have the Fourier transforms
\begin{equation}
\hat{\Delta}^{\pm}(k)= \int d^4x \, \Delta^{\pm}(x) e^{ikx}=
\pm 2 \pi i \,  \Theta(\pm k^0) \delta(k^2) , \label{dfourier}
\end{equation}
where $\Theta$ is the Heaviside distribution.
The fact that the commutator
\begin{equation}
[\Phi(x),\Phi(0)]=-i\Delta^+(x)-i\Delta^-(x) =: -i \Delta(x)
\end{equation}
vanishes for spacelike arguments (with $x^2 < 0$)
due to the requirement of microcausality, leads to
the important property that the Jordan-Pauli distribution
$\Delta$ has {\emph{causal support}}, i.e. it vanishes outside the closed
forward and backward light cone such that
\begin{equation}
\mbox{supp} \, \Delta(x) \subseteq \overline{V}^- \cup \overline{V}^+  \, , \quad
\overline{V}^\pm=\{x \, | \, x^2 \ge 0, \, \pm x^0 \ge 0 \}
\end{equation}
in the sense of distributions.
Furthermore, the Jordan-Pauli distribution has a simple representation
in configuration space
\begin{equation}
\Delta(x)=\frac{1}{2 \pi } \mbox{sgn}(x^0) \delta(x^2),
\end{equation}
and solves the wave equation $\Box \Delta(x) = 0$ with the Cauchy data
$\Delta(0,\vec{x})=0$ and $(\partial_0 \Delta)(0, \vec{x})= \delta^{(3)}(\vec{x})$.

A further crucial observation is the fact that one can introduce the
retarded propagator $\Delta^{ret}(x)$ which coincides with $\Delta(x)$ on
$\overline{V}^+ \! - \{0 \}$, i.e. $\Delta^{ret} (\varphi) =
\Delta (\varphi)$ holds for all test functions in the
Schwartz space $\varphi \! \in
\! \mathcal{S}(\mathds{R}^4)$ with support $\mbox{supp} ( \varphi ) \subset
\mathds{R}^4 - \overline{V}^-$.
In configuration space, $\Delta^{ret}(x)$ is obviously given by
\begin{equation}
\Delta^{ret}(x)=\frac{1}{2 \pi} \Theta (x^0) \delta(x^2),
\end{equation}
and as a special case of the edge of the wedge theorem \cite{PCT}
it is known that the Fourier transform of the retarded distribution
$\hat{\Delta}^{ret}(k)$ is the boundary value of an analytic function
$r(z)$, regular in $T^+ := \mathds{R}^4+ i V^+$. It is given by
\begin{equation}
\hat{\Delta}^{ret}(k)=-\frac{1}{k^2+ik^0 0}=-P\frac{1}{k^2}+i \pi \mbox{sgn}(k^0)
\delta(k^2).
\end{equation}
The analytic expression for the Feynman propagator is recovered from
the observation that
\begin{equation}
\Delta_F(x)=-\Delta^{ret}(x)+\Delta^-(x),
\end{equation}
since
\begin{equation}
\langle 0 | T ( \Phi(x) \Phi(0) ) | 0 \rangle = \Theta(x^0) [\Phi(x),\Phi(0)] +
\langle 0 | \Phi(0) \Phi(x) | 0 \rangle ,
\end{equation}
and consequently
\begin{equation}
\hat{\Delta}_F(k)=-\hat{\Delta}^{ret}(k)+\hat{\Delta}^-(k)=\frac{1}{k^2+i 0} .
\end{equation}
The retarded distribution $\Delta^{ret}(x)$ is a weak solution of the inhomogeneous wave
equation
\begin{equation}
\Box \Delta^{ret}(x) = \delta^{(4)} (x) = - \Box \Delta_F (x).
\end{equation}
Note that the Feynman propagator can also be written as
\begin{equation}
\Delta_F(x)=-\Theta (x^0) \Delta^+ (x) + \Theta(-x^0) \Delta^- (x). \label{freqdecomp}
\end{equation}

For later use we introduce now distributions $E_n(x)$, fulfilling the 
n-fold iterated wave equation $\Box^n E_n(x)=0$, with the properties
\begin{equation}
E_1(x)=\Delta(x),
\end{equation}
\begin{equation}
\Box E_{n+1}(x)=E_n(x),
\end{equation}
\begin{equation}
E_n(\lambda x)=\lambda^{2n-4} E_n(x), \quad \lambda \! \in \! \mathds{R}, \label{retscaling}
\end{equation}
satisfying the complete set of Cauchy data at $x^0=0$
\begin{equation}
(\partial_0^k E)(0,\vec{x})=0, \quad k=0,...,2n-1, \quad (\partial_0^{2n-1}E)(0,\vec{x})=
\delta^{(3)}(\vec{x}).
\end{equation}

The $E_n$'s can be constructed in a straightforward way. We consider first
the distribution
\begin{equation}
E(x)=\frac{1}{8 \pi} \mbox{sgn}(x^0) \Theta(x^2).
\end{equation}
From
\begin{equation}
\partial_\nu \Theta(x^2) = 2 x_\nu \Theta' (x^2)=2 x_\nu \delta(x^2),
\end{equation}
\begin{equation}
\partial^\nu (2 x_\nu \delta(x^2))=8 \delta(x^2) + 4 x^2 \delta' (x^2),
\end{equation}
one obtains by means of the identity ($x^2 \delta(x^2)=0$)
\begin{equation}
x^2 \delta'(x^2) = \frac{d}{d x^2} (x^2 \delta(x^2))-\delta(x^2) = - \delta(x^2)
\label{identity}
\end{equation}
\begin{equation}
\Box \Theta(x^2)=4 \delta(x^2)
\end{equation}
and it is a simple exercise to show that also
\begin{equation}
\Box (\mbox{sgn}(x^0) \Theta(x^2))=4 \mbox{sgn} (x^0) \delta(x^2)
\end{equation}
holds.
Therefore, one may define
\begin{equation}
E_2(x):=E(x)=\frac{1}{8 \pi} \mbox{sgn}(x^0) \Theta(x^2),
\end{equation}
since $\Box E(x)=\Delta(x)$ and $E_2$ has the correct scaling behavior
and fulfills also all other requirements given above.
From $\Box E_2(x) = \Delta(x)$, one obtains in momentum space
\begin{equation}
p^2 \hat{E}_2(k) = -\frac{i}{2 \pi} \mbox{sgn}(k^0) \delta(k^2)
\end{equation}
and by means of the identity eq. (\ref{identity}) with $x$ replaced by $k$
one may write
\begin{equation}
\hat{E}_2(k) = \frac{i}{2 \pi} \mbox{sgn} (k^0) \delta' (k^2).
\end{equation}
All these calculations are formal to a certain extent, but correct.
However, one should bear in mind that the meaning of the distribution $\delta'(k^2)$ is
rather defined by $\hat{E}_2$ given above than vice versa. 
Note that the positive frequency part
\begin{equation}
\mbox{`} \hat{E}^{\pm}_2(k) \mbox{'} = \frac{i}{2 \pi} \Theta(k^0) \delta' (k^2)
\end{equation}
is not unambiguously defined. Only derivatives of $E$ in configuration space can be split
unambiguously into positive and negative frequency parts, which emerge, e.g.,
in the photon propagator in the Landau gauge
\begin{equation}
\Delta_{Landau}^{\mu \nu} (k) \sim \frac{g^{\mu \nu}-\frac{k^\mu k^\nu}{k^2+i0}}{k^2+i0},
\end{equation}
in the term $\sim k^\mu k^\nu / k^4$.
We give here simply the regularized result for $E_2^{\pm}$ . One has
\begin{equation}
E^{\pm}_2 (x)=\mp \frac{i}{16 \pi^2}
\log \Biggl( - \frac{x^2 \mp i x^0 0}{\lambda_\mathcal{R}^2} \Biggr),
\end{equation}
or
\begin{displaymath}
E^{-}_2 (x)=+\frac{i}{16 \pi^2}
\log |x^2 / \lambda_\mathcal{R}^2 |
+\frac{1}{16 \pi} \mbox{sgn}(x^0) \Theta(x^2),
\end{displaymath}
\begin{displaymath}
E^{+}_2 (x)=-\frac{i}{16 \pi^2}
\log | x^2 / \lambda_\mathcal{R}^2 |
+\frac{1}{16 \pi} \mbox{sgn}(x^0) \Theta(x^2),
\end{displaymath}
where $\lambda_\mathcal{R}$ is a renormalization length scale, and combining
$E_2^+$ and $E_2^-$ indeed gives
\begin{equation}
E_2^+ (x)+E_2^- (x) =\frac{1}{8 \pi} \mbox{sgn} (x^0) \Theta(x^2).
\end{equation}
This result is directly related to the fact that the convolution
of the Feynman propagator in configuration space appearing in $S_3$
which corresponds to the formal expression $\sim (k^2+i0)^{-2}$ in momentum space
(see eq. \ref{Sn}) leads to an integral of the form
\begin{displaymath}
\Gamma_2^{\mathcal{R}}(k)= \mathcal{R} \Biggl[
\int \frac{d^4 x_1}{(2 \pi)^4} \frac{1}{x_1^2-i0} \, \frac{1}{(x-x_1)^2-i0}
\Biggr] \label{regula}
\end{displaymath}
\begin{equation}
=\frac{i}{4(2 \pi)^2} \log \Biggl (-\frac{x^2-i0}{\lambda_{\mathcal{R}}^2} \Biggr)=
\frac{i}{4(2 \pi)^2} \log|x^2/\lambda_{\mathcal{R}}^2|-
\frac{1}{16 \pi} \Theta(x^2) \, , \label{gamma2}
\end{equation}
when the adiabatic limit in the variable $x_1$ is performed.
$\mathcal{R}$ denotes the regularization procedure with renormalization
scale $\lambda_{\mathcal{R}}$. Obviously, the regularized expression
for $\Gamma_2^{\mathcal{R}}(x)$ is defined up to a constant
in real space or up to a local distribution in momentum space $\sim \delta^{(4)}(k)$.
The scaling symmetry of the Feynman propagator $\Delta_F(\lambda x)=\lambda^{-2}
\Delta_F(x)$, $\lambda \! \in \! \mathds{R}$,
is {\em{spontaneously broken}} by regularization, such that a corresponding
scaling law $\Gamma_2^{\mathcal{R}}(\lambda x) = \Gamma_2^{\mathcal{R}}(x)$
does {\it not} hold as one might expect naively from the formal definition
of $\Gamma_2(x)$ in eq. (\ref{Sn}). The adiabatic limit {\emph{does not}} exist
for the $S_n$ for $n \ge 3$ for the C-number distributions.

Finally, from
\begin{equation}
\Box (x^2)^n = 4 n (n+1) (x^2)^{n-1},
\end{equation}
we immediately deduce
\begin{equation}
E_n(x)=\frac{1}{(2 \pi) 4^{n-1} (n-1)! (n-2)!} (x^2)^{n-2} \mbox{sgn}(x^0) \Theta (x^2), \quad
n \ge 2.
\end{equation}

\section{Yang-Feldman equations}
Since the $E_n$'s introduced in the previous section all have causal support,
we can define retarded distributions
\begin{equation}
E_n^{ret}(x)=\frac{1}{(2 \pi) 4^{n-1} (n-1)! (n-2)!} (x^2)^{n-2} \Theta (x^0)
\Theta (x^2), \quad n \ge 2,
\end{equation}
fulfilling $\Box^n E_n^{ret} (x) = \delta^{(4)} (x)$.
Roughly speaking, the distribution $E_n^{ret}$ corresponds to the expression
$(-1)^{n}(k^2+i k^0 0)^{-n}$ in momentum space (the minus signs stem from
the fact that the wave operator $\Box$ corresponds to $-k^2$ in momentum
space). Note that the $E_n^{ret}$ still respect the scaling law eq. (\ref{retscaling}).
Performing now a resummation of the distributions in configuration space
corresponding to minus
\begin{equation}
\frac{1}{k^2+i k^0 0} + \frac{m^2}{(k^2+i k^0 0)^2}+ \frac{m^4}{(k^2+i k^0 0)^3}+...
\label{retsum}
\end{equation}
instead of minus
\begin{equation}
\frac{1}{k^2+i0} + \frac{m^2}{(k^2+i0)^2}+ \frac{m^4}{(k^2+i0)^3}+... ,
\end{equation}
leads to
\begin{displaymath}
\sum \limits_{n=1}^{\infty} (-1)^{n-1} (m^2)^{n-1} E_n^{ret}(x)
\end{displaymath}
\begin{displaymath}
=\Delta^{ret}(x)+\frac{1}{(2 \pi)}
\sum \limits_{n=2}^{\infty}
\frac{(-m^2)^{n-1}}{4^{n-1} (n-1)! (n-2)!} (x^2)^{n-2} \Theta (x^0) \Theta(x^2)
\end{displaymath}
\begin{equation}
=\frac{1}{2 \pi} \Theta (x^0) \Biggl[
\delta(x^2)-\Theta(x^2) \frac{m}{2 \sqrt{x^2}} J_1 \Bigl( m \sqrt{x^2} \Bigr) \Biggr],
\label{massivepropagator}
\end{equation}
where $J_1$ is the Bessel function of order 1. The result in eq. (\ref{massivepropagator})
is indeed the massive retarded propagator $\Delta^{ret}_m$, i.e. the Fourier transform
of $-(k^2-m^2+ik^0 0 )^{-1}$ !

A way to introduce the resummed retarded distribution constructed above
is supplied by the Yang-Feldman formalism.
One may write the interacting scalar field as a formal power series
$\Phi_{int}=\sum_{n=0}^{\infty} (m^2)^n \Phi_n$, that fulfills the equation of
motion
\begin{equation}
\Box \Phi_{int} = -m^2 g \Phi_{int},
\end{equation}
where $\Phi_0$ is the free massless (incoming) field.
 The higher terms are recursively defined by
\begin{equation}
\Phi_n(x)= - \int d^4 x_1 \Delta^{ret} (x-x_1) g(x_1) \Phi_{n-1} (x_1). \label{recursion}
\end{equation}
Without the test function $g$, the integral eq. (\ref{recursion}) would be meaningless
even if one chooses a massive scalar field for $\Phi_0$.
For example, the expression
\begin{displaymath}
\int d^4 x_1 \Delta^{ret}(x-x_1) \Phi(x_1)g(x_1)=
\end{displaymath}
\begin{equation}
\frac{1}{4 \pi} \int d^3 \vec{x}_1 \frac{g(x^0-|\vec{x}-\vec{x}_1|,\vec{x}_1)}
{|\vec{x}-\vec{x}_1|} \Phi (x^0-|\vec{x}-\vec{x}_1|,\vec{x}_1) \label{divop}
\end{equation}
becomes meaningless in the limit $g \rightarrow 1$ and contains a volume divergence,
and the situation is similar in the massive case. Such issues are rarely discussed in
the literature, and completely ignored in the original work of Yang and Feldman
\cite{YangFeldman}. Some attempts to deal in a rigorous style with the problem of
the adiabatic limit can be found in \cite{GScharf,DutschFredenhagen,DoscherZahn}.

The inductive construction of the retarded distributions can also be understood
from the following simple calculation in configuration space.
Convolving $\Delta^{ret}$ with an $E^{ret}_n$ leads to an integral for $n \ge 2$
\begin{displaymath}
I(x)=\frac{(2 \pi)^{-2} 4^{1-n}}{(n-1)! (n-2)!} \times
\end{displaymath}
\begin{equation}
\int d^4 x_1 \Theta(x_1^0) \delta(x_1^2) ((x-x_1)^2)^{n-2} \Theta (x^0-x_1^0)
\Theta ((x-x_1)^2). \label{convolution}
\end{equation}
The integral $I$ vanishes outside the closed forward light cone due to Lorentz invariance
and the two $\Theta$-distributions in eq. (\ref{convolution}). Therefore we can go
to a Lorentz frame where $x=(x^0>0, \vec{0})$ and $(x-x_1)^2=(x^0)^2-2x^0 |\vec{x}_1|$ due
to the $\delta$-distribution in eq. (\ref{convolution}), and we obtain
\begin{displaymath}
I(x^0>0,\vec{0})=\frac{(2 \pi)^{-2} 4^{1-n}}{(n-1)! (n-2)!} \times
\end{displaymath}
\begin{displaymath}
\int \frac{d^3 \vec{x}_1}{2 |\vec{x}_1|} ((x_0-x_1)^2)^{n-2}
\Theta((x^0)^2-2x^0 |\vec{x}_1|) \Theta(x^0-|\vec{x}_1|)
\end{displaymath}
\begin{displaymath}
=\frac{(2 \pi)^{-1} 4^{1-n}}{(n-1)! (n-2)!}
\int d|\vec{x}_1| |\vec{x}_1| (((x^0)^2-2 x_0 |\vec{x}_1|)^2)^{n-2} \Theta
(x^0/2-|\vec{x}_1| )
\end{displaymath}
\begin{equation}
=\frac{(2 \pi)^{-1} 4^{1-n}}{(n-1)! (n-2)!} \times \frac{((x^0)^2)^{n-1}}{4 n (n-1)},
\end{equation}
and the Lorentz invariant expression for $I$ becomes
\begin{equation}
I(x)=\frac{1}{(2 \pi) 4^n n! (n-1)!} (x^2)^{n-1} \Theta(x^0) \Theta(x^2)=E_{n+1}(x),
\end{equation}
i.e. we recover $E_{n+1}^{ret}$ as defined above. This shows that a formalism using
retarded products of distributions has an advantage compared to the
common strategy to work with time-ordered products. It is clear that the adiabatic
limit for the final integral eq. (\ref{divop}) defining the Yang-Feldman field operator
cannot be performed, however, the adiabatic limit exists for the full recursively defined
retarded propagator and this allows to reconstruct the massive theory.
Of course, according to Haag's theorem \cite{Haag},
representations of the canonical commutation relations algebra
to different masses are inequivalent, and an attempt to express the massive field
operator by the massless field would be futile.

\section{Relation to the operator product expansion}
In this section, we briefly describe the relation of our findings
to a well-known technique, namely the operator product expansion (OPE),
which goes back to a work of Wilson \cite{Wilson}. The OPE
provides a method to expand singular operator products
as a sum of nonsingular local operators with coefficients
being (singular) C-number distributions.
The problem with the series eq. (\ref{summation}) which is solved
within the framework presented above can be related to the fact that
only quantities defined at short distances compared to the characteristic
length of the physical system under consideration, i.e. short compared to
the length scale $1/m$ in our case, can be expanded in powers of $m^2$.

At short space-time distances, one may consider the time-ordered product
of two scalar fields and expand the corresponding vacuum expectation value
\begin{equation}
\langle 0 | T (\Phi(x) \Phi(0)) | 0 \rangle = C_1(x^2)+
C_{m^2}(x^2) \langle 0 | [\Phi^2]  | 0 \rangle
\end{equation}
with Wilson coefficients $C_1$ and $C_{m^2}$, where
$\langle 0 | [\Phi^2] | 0 \rangle$ is the vacuum expectation value
in the free massive theory of the regularized composite
operator
\begin{equation}
\langle 0 | [\Phi^2] | 0 \rangle=\frac{m^2}{16 \pi^2}
\log ( m^2/\mu_\mathcal{R}^2 )
\end{equation}
with a renormalization scale $\mu_\mathcal{R}$ \cite{Brown}.
Clearly, the whole propagator cannot be expanded in powers of $m^2$
due to the logarithmic dependence of $\langle 0 | [\Phi^2] | 0 \rangle$.
However, the Wilson coefficients can be expanded in powers of $m^2$, and
the corresponding power series are even convergent. An explicit calculation,
performed in Euclidean space as it is common in the literature, leads to
($r^2 \sim -x^2$)
\begin{equation}
C_1(r^2)=\frac{1}{4 \pi^2 r^2} \bigl( 1+ mr I_1(mr) \log(\mu_\mathcal{R} r/2)
+ \, {\mbox{even powers in}} \, \, mr \bigr), \label{Wilson1}
\end{equation}
\begin{equation}
C_{m^2}(r^2)=\frac{2 I_1(mr)}{mr},
\end{equation}
where $I_1$ is the modified Bessel function of order 1.
Note that the $\mu_\mathcal{R}$-dependence of $C_1(r^2)$ is canceled by that of
$\langle 0 | [\Phi^2] | 0 \rangle$.

The correction to the commutator or retarded propagator originates from
the cut of $\log ( \mu_\mathcal{R} r)$ in the complex plane of $r^2$.
From  eq. (\ref{Wilson1}) one obtains
\begin{equation}
\pi \frac{1}{4 \pi^2 r^2} mr I_1(mr)=\frac{m}{4 \pi \sqrt{r^2}}
I_1(m \sqrt{r^2}), \label{cut}
\end{equation}
and returning to Minkowski space by further using the identity
$I_1(z)=\frac{1}{i} J_1(iz)$,
eq. (\ref{cut}) is in perfect agreement with the corresponding Bessel term
in eq. (\ref{massivepropagator}).
The problem of the series eq. (\ref{summation}) is related to the
non-trivial $m^2$-dependence of the local composite operator vacuum
expectation value $\langle 0 | [\Phi^2] | 0 \rangle$ in the framework of
the OPE.

\section{Mass terms for gauge fields}
It is an interesting question to what extent one could describe theories
with interacting massive vector bosons on the basis of massless gauge
fields. To make a step towards this direction, we consider a pure
SU(N) Yang-Mills theory without fermions below,
and call the massless gauge bosons gluons for short in the forthcoming.

As a first order coupling, one can choose
\begin{equation}
T_1(x)=ig_{_{YM}} f_{abc} \{\frac{1}{2}
: \! A_{\mu a}(x) A_{\nu b}(x) F^{\nu \mu}_c (x) \! :
-: \! A_{\mu a}(x) u_b (x) \partial^{\mu} {\tilde{u}}_c(x) \! : \} , \label{QCD}
\end{equation}
where $F_a^{\nu \mu} = \partial^\nu A^\nu_a - \partial^\mu A^\nu_a$ is
the free field strength tensor, $u_a, \tilde{u}_a$ are the
(fermionic) ghost fields, and the $f_{abc}$ are the
SU(N) structure constants corresponding to the color indices $a$,$b$ and $c$.
This first order coupling contains the well-known three-gluon vertex
and a ghost-antighost-gluon coupling. A commutator term $\sim
g f_{abc} A^\nu_b A^\mu_c$ ist absent in the free expression for $F_a^{\nu \mu}$,
since this term contains an additional Yang-Mills coupling constant $g_{_{YM}}$, and the
resulting four-gluon coupling term generated by such an additional non-abelian
term would be of the order $g_{_{YM}}^2$.

In the 't Hooft-Feynman gauge, the asymptotic massless free fields in eq. (\ref{QCD})
satisfy the commutation relations \cite{DutschSinv}
\begin{equation}
[A_{\mu a}^{(\pm)}(x),A_{\nu b}^{(\mp)}(y)]=i \delta_{ab}
g_{\mu \nu} \Delta^{\mp}(x-y)
\end{equation}
and
\begin{equation}
\{u_a^{(\pm)}(x),\tilde{u}_b^{(\mp)}(y)\}=-i \delta_{ab} \Delta^{\mp}(x-y),
\end{equation}
and all other \{anti-\}commutators vanish.
It can be shown that the coupling eq. (\ref{QCD}) respects perturbative gauge invariance
which corresponds to the full BRST symmetry of the theory \cite{BRST,Tyutin}.
The introduction of ghost couplings is necessary to preserve perturbative quantum
gauge invariance at first order and unitarity at second order in $g_{_{YM}}$.
We give here a short
definition of perturbative BRST symmetry based on a gauge charge operator $Q$
(for details refer to \cite{YM,YMII}).

Defining the gauge charge $Q$, which is the generator of gauge transformations,
by
\begin{equation}
Q:=\int d^3x \partial_\mu A^\mu_a (x)  \stackrel
{\leftrightarrow}{\partial_0} u_a(x) ,
\end{equation}
where summation over repeated indices is understood,
one is led to the following (anti-) commutators with the fields:
\begin{equation}
[Q,A_a^\mu]=i\partial^\mu u_a ,
\quad [Q,F_a^{\mu \nu}]=0  , \quad
\{Q,u_a\}=0  , \quad \{Q,\tilde{u}_a\}= -i\partial_\mu A^\mu_a .
\label{comm}
\end{equation}
The operator $Q$ has been introduced for the first time by Kugo and Ojima \cite{Kugo}.
The pseudounitary gauge transformation of a field operator like, e.g., $A_\mu$, is
implemented by
\begin{equation}
A'_\mu = e^{-i \lambda Q} A_\mu e^{i \lambda Q}, \quad \lambda \! \in \! \mathds{R}.
\end{equation}
Calculating $[Q,T_1(x)]$ leads after some algebra to the interesting result
\begin{equation}
[Q,T_1]=i \partial_\mu T^\mu_{1/1},
\end{equation}
\begin{equation}
T^\mu_{1/1} = ig f_{abc}\bigl\{ : \! u_a A_{\nu b} F^{\mu \nu}_c \! :
+ \frac{1}{2} : \! u_a u_b \partial^\mu \tilde{u}_c \! : \}.
\end{equation}
The first order coupling is gauge invariant through the presence of
a ghost coupling term in $T_1$ in the sense that the commutator of $Q$ with
$T_1$ is a pure divergence in the the analytic sense, i.e. an infinitesimal
gauge transformation adds only a divergence to the original interaction term.

This actual definition of perturbative gauge invariance can be generalized to
higher orders of perturbation theory by the requirement that \cite{YM,YMII}
\begin{equation}
[Q,T_n(x_1,...x_n)] = i \sum_{l=1}^{n} \partial_\mu^{x_l} T_{n/l}^{\mu}
(x_1,...x_n) = (\hbox{\rm sum of divergences}) , \label{dive}
\end{equation}
where $T^\mu_{n/l}$ is a mathematically rigorous (regularized) version of the
time-ordered product
\begin{equation}
T^\mu_{n/l}(x_1,...,x_n)
\, \mbox{`} \! = \! \mbox{'} \, \, T(T_1(x_1)...T^\mu_{1/1} (x_l)...T_1(x_n)). \label{Slavnov}
\end{equation}
The identities eq. (\ref{Slavnov}) imply the well-known Slavnov-Taylor
identities, which express gauge symmetry on an analytic level for
Greens functions \cite{DutschSlavnov}.

As aforementioned, the four-gluon term $\sim g_{_{YM}}^2$
is missing in $T_1$. This term appears in a natural way
as a necessary local normalization term of the gluon-gluon
scattering diagram at second order in order to preserve
gauge invariance in the present framework.
$Q$ has the important property that it is nilpotent, i.e. $Q^2=0$.
This basic property of $Q$ and the so-called Krein structure on the
Fock-Hilbert space \cite{Razumov,Bognar} allows to prove unitarity of the
$S$-matrix on the physical Hilbert space $\mathcal{H}_{phys}$, which is a
subspace of the Fock-Hilbert space $\mathcal{F}$ containing also the
unphysical ghosts and unphysical degrees of freedom of the vector fields.

The basic question is now whether one can add a mass term to the
interaction, such that the gauge symmetry remains preserved in a
certain `hidden' way.
Adding simply a Proca mass term to $T_1$
\begin{equation}
T_1^{m,A} = \frac{i}{2} m_a^2 : \! A_\mu^a A^\mu_a \! :
\end{equation}
would destroy gauge invariance at first order, since
\begin{equation}
[Q, : \! A_\mu^a A^\mu_a \! : ]=2i  : \! \partial_\mu u_a A^\mu_a \! :
\end{equation}
cannot be written as a divergence. But this defect can be remedied
by adding additionally a ghost mass term 
\begin{equation}
T_1^{m,u}= i m_a^2 : \! u_a \tilde{u}_a \! : =  -i m_a^2 : \! \tilde{u}_a u_a \! :
\end{equation}
to $T_1$, since
\begin{equation}
[Q, : \! u_a \tilde{u}_a \! :]= : \! \{ Q, u_a \} \tilde{u}_a \! : -
: \! u_a \{ Q, \tilde{u}_a \} \! : =
i : \! u_a \partial_\mu A^\mu_a \! :,
\end{equation}
and so
\begin{equation}
[Q,T_1^{m,A}+T_1^{m,u}]= - m_a^2 (: \! \partial_\mu u_a A^\mu_a \! : +
: \! u_a \partial_\mu A^\mu_a \! :) = - m_a^2 \partial_\mu : \! 
u_a A^\mu_a \! : ,
\end{equation}
such that $T_1^{m,A}+T_1^{m,u} $ is a gauge invariant mass term at first order in
$m^2$. The vector boson and ghost mass
is equal for a certain fixed color in the 't Hooft-Feynman gauge, and an analogous
result holds for arbitrary renormalizable $R_\xi$ gauges. It is clear that a resummation
of the mass term will change the theory completely, and massive longitudinal
gluon modes will become physical states. This will necessitate the introduction of
new physical and unphysical Higgs (ghost) fields and an adjustment of the boson masses
in the theory according to the symmetry breaking scenario, in order to save the
consistency of the new theory. The full story is therefore not so simple \cite{Electroweak}.
There original gauge symmetry of the massless theory leaves its traces in
the massive theory, but one should not expect that is it possible to construct
a Higgs free electroweak model of interacting massive vector bosons based on the
considerations presented above. However, it should be pointed out that the widespread
statement in the literature that the Higgs mechanism is responsible for the particle
masses is misleading to some extent. One could also argue that the presence of Higgs fields
is necessary under certain circumstances in order to save the consistency of the
theory, when vector bosons become massive.

\section{Concluding remark}
The series eq. (\ref{retsum}) converges formally to the massive retarded propagator,
but not in the distributional sense, i.e. not in the $\mbox{weak}^\star$ topology.
This can be realized in a simple manner
by the observation that every individual term in eq. (\ref{retsum}) displays
a singular behavior on the light cone, whereas the massive propagator is singular
on the mass shell. It is instructive to compare the situation to the following simple
example in one dimension. We consider distributions
\begin{equation}
d_n (x) := \sum \limits_{k=0}^{n} \frac{(-1)^k}{k!} \delta^{(k)} (x),
\end{equation}
where $\delta^{(k)} (x)$ denotes the $k$-th derivative of the one-dimensional
Dirac distribution here.
Applying $d_n$ to a test function $\varphi(x) \! \in \! \mathcal{S}(\mathds{R})$
results in
\begin{equation}
d_n(\varphi)=\sum \limits_{k=0}^{n} \frac{\varphi^{(k)}(0)}{k!}, \label{translation}
\end{equation}
and all $d_n$ have local support $\mbox{supp} (d_n) = \{ 0 \}$.
If the test function can be continued to a holomorphic
function in an open region containing the real axis (i.e. $\varphi$ is analytic in the
sense of $\varphi \! \in \! \mathcal{S}(\mathds{R}) \cap C^\omega(\mathds{R})$,
and not only a rapidly decreasing smooth $C^\infty(\mathds{R})$ function),
then the Taylor expansion eq. (\ref{translation}) indeed converges to
\begin{equation}
\lim_{n \to \infty} d_n (\varphi)= \sum \limits_{k=0}^{\infty}
\frac{\varphi^{(k)}(0)}{k!} =\varphi(1)
\end{equation}
and for the restricted space of test functions
$\mathcal{S}(\mathds{R}) \cap C^\omega(\mathds{R})$
one obtains a weak limit
\begin{equation}
\lim_{n \to \infty} d_n=
\sum \limits_{k=0}^{\infty} \frac{(-1)^k}{k!} \delta^{(k)} (x) = \delta(x-1),
\quad \mbox{supp} ( \delta(x-1))=\{ 1 \}.
\end{equation}
This suggests the speculative question whether the choice of a modified space of test
functions would have advantages compared to $\mathcal{S}(\mathds{R}^4)$ concerning
the consistency of the mathematical framework of non-perturbative quantum field theory.
Some ideas to generalize the admissible test function spaces in quantum field
theory have been investigated by Jaffe \cite{Jaffe}. The idea to construct a
`hyperfunction quantum field theory' has been followed in
\cite{Nagamachi,Bruning}. There, the main technical problem is the fact that
rapidly decreasing holomorphic functions do not have compact support, such that
one has to look for a new concept to express the fundamental causality condition which
is expressed in the Wightman formalism by eq. (\ref{causality}).

\section*{Acknowledgements}
This work was supported by the Swiss National Science Foundation.


\section*{References}
\begin{thebibliography}{99}

\bibitem{Constantinescu}
Constantinescu, F.:
{\it Distributions and their applications in physics}.
Pergamon Press, Oxford (1980).

\bibitem{Schw1}
Schwinger, J.:
On gauge invariance and vacuum polarization.
Phys. Rev. {\bf 82}, 664 (1951).

\bibitem{ASchw}
Aste, A., Scharf, G., Walther, U.:
Power counting degree versus singular order in the Schwinger model.
Nuovo Cim. {\bf A111}, 323 (1998).

\bibitem{PCT}
Streater, R.F., Wightman, A.S.:
{\it PCT, Spin, statistics and all that}.
Benjamin-Cummings Publishing Company, New York (1964).

\bibitem{GScharf} Scharf, G.: {\it Finite quantum electrodynamics}.
2nd ed., Springer Verlag, New York (1995).

\bibitem{Aste} Aste, A.:
The two loop master diagram in the causal approach.
Annals Phys. {\bf 257}, 158 (1997).

\bibitem{AsteSunrise} Aste, A., Trautmann, D.:
Finite calculation of divergent self-energy diagrams.
Can. J. Phys. {\bf{81}}, 1433 (2003).

\bibitem{eg}
Epstein H., Glaser V.: 
The role of locality in perturbation theory.
Annales Poincar\'{e} Phys. Theor. {\bf{A19}}, 211 (1973).

\bibitem{Coleman}
Coleman, S.:
There are no Goldstone bosons in two dimensions.
Commun. Math. Phys. {\bf{31}}, 259 (1973).

\bibitem{Positivity}
Wightman, A.S.: {\it Cargese lectures of theoretical physics}.
Gordon and Breach, New York (1964).

\bibitem{Abdalla}
Abdalla, E., Abdalla, M., Rothe, K.:
{\it 2-dimensional quantum field theory}.
World Scientific, London (1991).

\bibitem{YangFeldman}
Yang, C.N., Feldman, D.:
The S-matrix in the Heisenberg representation.
Phys. Rev. {\bf{79}}, 972 (1950).

\bibitem{DutschFredenhagen}
D\"utsch, M., Fredenhagen, K.:
Causal perturbation theory in terms of retarded products, and a
proof of the action Ward identity.
Rev. Math. Phys. {\bf{16}}, 1291-1348 (2004).

\bibitem{DoscherZahn}
D\"oscher, C., Zahn, J.:
Infrared cutoffs and the adiabatic limit in non-commutative space-time.
Phys. Rev. {\bf{D73}}, 045024 (2006).

\bibitem{Haag}
Haag, R.:
On quantum field theories.
Dan. Mat. Fys. Medd. {\bf{29}}, no. 12, 1 (1955).

\bibitem{Wilson}
Wilson, K.G.:
Non-Lagrangian models of current algebra.
Phys. Rev. {\bf{179}}, 1499-1512 (1969).

\bibitem{Brown}
Brown, L.S.: {\it Quantum field theory}.
Cambridge University Press, Cambridge (1992).

\bibitem{DutschSinv}
Aste, A., Scharf, G., D\"utsch, M.:
Gauge independence of the S-matrix in the causal approach.
J. Phys. {\bf{A31}}, 1563 (1998).

\bibitem{BRST}
Becchi, C.,Rouet, A., Stora, R.:
Renormalization of gauge theories.
Ann. Phys. {\bf{98}}, 287 (1976).

\bibitem{Tyutin}
Tyutin, I.V.: Lebedev Report No. FIAN No. 39, 1975 (unpublished).

\bibitem{Kugo}
Kugo, T., Ojima, I.:
Local covariant operator formalism of non-abelian gauge theories and quark
confinement problem.
Suppl. Prog. Theor. Phys. {\bf {66}}, 1 (1979).

\bibitem{YM}
Aste A., Scharf, G.:
Non-abelian gauge theories as a consequence of perturbative quantum gauge invariance.
Int. J. Mod. Phys. {\bf{A14}}, 3421 (1999).

\bibitem{YMII}
D\"utsch, M., Hurth, T., Krahe, F., Scharf, G.:
Causal construction of Yang-Mills theories: II.
Nuovo Cim. {\bf{A107}}, 375 (1994).

\bibitem{DutschSlavnov}
D\"utsch, M.:
Slavnov-Taylor Identities from the Causal Point of View.
Int. J. Mod. Phys. {\bf{A12}}, 3205 (1997).

\bibitem{Razumov}
Razumov, A.V., Rybkin, G.N.:
State space in BRST-quantization of gauge-invariant systems.
Nucl. Phys. {\bf {B332}}, 209 (1990).

\bibitem{Bognar}
Bognar, J.:
{\em {Indefinite inner product spaces}}.
Springer, Berlin (1974).

\bibitem{Electroweak}
Aste, A., Scharf, G., D\"utsch, M.:
Perturbative gauge invariance: Electroweak theory II.
Annalen Phys. {\bf{8}}, 389 (1999). 

\bibitem{Jaffe}
Jaffe, A.:
High-energy behavior in quantum field theory. I. Strictly localizable fields.
Phys. Rev. {\bf{158}}, 1454 (1967).

\bibitem{Nagamachi}
Nagamachi, S., Mugibayashi, N.:
Hyperfunction quantum field theory.
Commun. Math. Phys. {\bf{46}}, 119 (1976).

\bibitem{Bruning}
Br\"uning, E., Nagamachi, S.:
Hyperfunction quantum field theory: Localized fields without localized test functions.
Lett. Math. Phys. {\bf{63}}, 141 (2003).

\end{thebibliography}
\end{document}